\documentclass[10pt,a4paper]{article}

\usepackage{graphicx}
\usepackage[linesnumbered,tworuled,vlined]{algorithm2e}
\usepackage{amsmath}
\usepackage{amssymb,theorem}

\DeclareMathAlphabet{\mathonebb}{U}{bbold}{m}{n}
\newcommand{\one}{\ensuremath{\mathonebb{1}}}

\usepackage{pgf,pgfarrows,pgfnodes,tikz,color,tkz-berge}
\usetikzlibrary{arrows,shapes,snakes,automata,backgrounds,petri,topaths,calc}

\DeclareMathOperator*{\argmin}{\arg\!\min}

\newcommand{\vect}[1]{\ensuremath{  #1 } }

\newcommand{\R}{{\mathbb R}}

\newcommand{\D}[2]{ \ensuremath{ \frac{\mathrm{d} #1 }{\mathrm{d} #2 } }}

\newcommand{\Pred}{\mathrm{Pred}}
\newcommand{\Succ}{\mathrm{Succ}}

\newtheorem{theorem}{Theorem}[section]
\newtheorem{lemma}[theorem]{Lemma}
\newtheorem{proposition}[theorem]{Proposition}
\theoremstyle{definition}

\theoremstyle{remark}
\newtheorem{remark}[theorem]{Remark}

\makeatletter
\newcommand*{\inlineequation}[2][]{%
  \begingroup
    \refstepcounter{equation}%
    \ifx\\#1\\%
    \else
      \label{#1}%
    \fi
    \relpenalty=10000 %
    \binoppenalty=10000 %
    \ensuremath{%
      #2%
    }%
    ~\@eqnnum
  \endgroup
}
\makeatother

\begin{document}

\title{Symbolic dynamics of biochemical pathways as finite states machines}

\author{Ovidiu Radulescu$^1$, Satya Swarup Samal$^2$, Aur{\'e}lien Naldi$^1$, \\
 Dima Grigoriev$^3$, and  Andreas Weber$^2$ \\
\small  $^1$ DIMNP UMR CNRS 5235, University of Montpellier 2, Montpellier, France, \\
\small  $^2$ Institut f\"{u}r Informatik II, Universit\"{a}t Bonn, Germany, \\
\small  $^3$ CNRS, Math\'ematiques, Universit\'e de Lille, 59655, Villeneuve d'Ascq, France.
 }

\maketitle

\abovedisplayskip=3pt
\belowdisplayskip=3 pt
\abovedisplayshortskip=0pt
\belowdisplayshortskip=0pt

\sloppy

\centerline{\bf Abstract}

We discuss the symbolic dynamics of biochemical networks with separate timescales. We show that symbolic dynamics of monomolecular
reaction networks with separated rate constants can be described by deterministic, acyclic automata with a number of states that is
inferior to the number of biochemical species. For nonlinear pathways, we propose a general approach to approximate their dynamics
by finite state machines working on the metastable states of the network (long life states where the system
has slow dynamics). For networks with polynomial rate functions we propose to compute metastable states as solutions of
the tropical equilibration problem. Tropical equilibrations are defined by the equality of at least two dominant monomials of
opposite signs in the differential equations of each dynamic variable. In algebraic geometry, tropical equilibrations are
tantamount to tropical prevarieties, that are finite intersections of tropical hypersurfaces.

\section{Introduction}
Networks of biochemical reactions are used in computational biology
as models of signaling, metabolism, and gene regulation.
For various applications it is important to understand how the dynamics of these models depend
on internal parameters and environment variables.
Traditionally, the dynamics of biochemical networks is studied in the framework of chemical kinetics
that can be either deterministic (ordinary differential equations) or stochastic (continuous time
Markov processes). Within this framework, problems such as
causality, reachability, temporal logics, are hard to solve and even to formalize.
Concurrency models such as Petri nets and process algebra conveniently formalize these questions
that remain nevertheless difficult.  The main source of difficulty is the
extensiveness of the set
of trajectories that have to be analysed. Discretisation of the phase space
does not solve the problem, because in multi-valued networks with $m$ levels (Boolean networks
correspond to $m=2$) the number of the states is $m^n$ and grows exponentially with the
number of variables $n$.
An interesting alternative to these approaches is symbolic dynamics which means replacing
the trajectories of the smooth system with a sequence of symbols. In certain cases, this
could lead to relatively simple descriptions. According to the famous
conjecture of Jacob Palis \cite{palis2000global},
smooth dynamical systems on compact spaces should have
a finite number of attractors whose basins cover the entire ambient space.
Compactness of ambient space  is satisfied by networks of biochemical reactions because of
conservation, or dissipativity.
For high dimensional systems with multiple separated timescales it it reasonable to
consider the following property: trajectories within basins of attraction
consists in a succession of fast transitions between relatively slow regions.
The slow regions, generally called metastable states,
can be of several types such as attractive invariant
manifolds, Milnor attractors or saddles.
Because of compactness of the ambient
space and smoothness of the vector fields defining the dynamics,
there should be a finite number of such metastable states.
This phenomenon,
called itinerancy received particular attention
in neuroscience \cite{tsuda1991chaotic}. We believe that similar phenomena occur in molecular
regulatory networks. A simple example is the set of bifurcations of metastable states
guiding the orderly progression of the cell cycle.
In this paper we use tropical geometry
methods to detect the presence of metastable states and describe
the symbolic dynamics as a finite state automaton.
The structure of the paper is the following.
In the second section we compute the symbolic dynamics of monomolecular
networks with totally separated constants. To this aim we rely on
previous results \cite{gorban-dynamic,radulescu2008robust,radulescu2012frontiers}.
In the third section we introduce tropical equilibrations of nonlinear networks.
Tropical equilibrations are good candidates for metastable states.
More precisely, we use minimal branches of tropical equilibrations as proxys for
metastable states.
In the forth section we propose an algorithm to learn
finite state automata defined on these states.
\section{Monomolecular networks with totally separated constants} \label{sec:mono}
Monomolecular reaction networks are the simplest reactions
networks. The structure of these networks is completely defined by a digraph $G=(V,{\mathcal A})$,
in which vertices $i \in V, 1 \leq i \leq n$ correspond to chemical species $A_i$,
edges $(i,j)\in {\mathcal A}$ correspond to
reactions $A_i \to A_j$ with kinetic constants $k_{ji} > 0$.
For each vertex, $A_i$, a positive real variable $c_i$ (concentration)
is defined.
The chemical kinetic dynamics is described by a system of linear differential equations
\begin{equation}
\D{c_i}{ t}=\sum_{j } k_{ij} c_j - (\sum_{j }
k_{ji})c_i ,
\label{kinur}
\end{equation}
where $k_{ji} > 0$ are kinetic coefficients. In matrix form one has : $\dot{c} = Kc$.
The solutions of \eqref{kinur} can be expressed in terms of left
and right eigenvectors of the kinetic matrix $K$:
\begin{equation} \label{solkinur}
\vect{c}(t)= \vect{r}^0 (\vect{l}^0, \vect{c}(0)) + \sum_{k=1}^{n-1} \vect{r}^k (\vect{l}^k, \vect{c}(0))
\exp(\lambda_k t) ,
\end{equation}
where $\vect{r}^k$, $\vect{l}^k$ are right and left eigenvectors of $\vect{K}$,
 $\vect{K}  \vect{r}^k = \lambda_k \vect{r}^k$,  $\vect{l}^k K   = \lambda_k \vect{l}^k$.

The system \eqref{kinur} has a conservation law $\D{}{t}(c_1 + c_2 + \ldots + c_n) = 0$,
and therefore there is a zero eigenvalue $\lambda_0=0$, $\vect{l}^0=(1,1,\ldots,1)$,
$(\vect{l}^0, \vect{c}(0)) = c_1(0) + c_2(0) + \ldots + c_n(0)$.
We say that the network constants are totally separated if for all $(i,j)\neq (i',j')$
one of the relations $k_{ji} \ll  k_{j'i'}$, or $k_{ji} \gg k_{j'i'}$ is satisfied.

It was shown in \cite{gorban-dynamic,radulescu2008robust,radulescu2012frontiers} that
the eigenvalues and the eigenvectors of an arbitrary monomolecular reaction
networks with totally separated constants can be approximated with good
accuracy by the eigenvalues of and the eigenvectors of a
reduced monomolecular networks whose reaction
digraph is acyclic (has no cycles), and deterministic
(has no nodes from which leave more than
one edge).
Let us denote by $G_r=(V_r,{\mathcal A}_r)$ the reduced digraph, and by $\kappa_i$ the
kinetic constant of the unique reaction that leaves a node $i \in V_r$.
The algorithm to obtain $G$ from $G_r$ can be found in
\cite{gorban-dynamic,radulescu2008robust,radulescu2012frontiers} and will not be repeated here.
Because $G_r$ is deterministic it defines a flow (discrete dynamical system) on the graph:
$\Phi(i)=j$, where $j$ is the unique node following $i$ on the digraph.
Reciprocally, we define $\Pred(i) = \phi^{-1}(i)$ as
the set of predecessors of the node $i$ in the digraph $G_r$,
namely $\Pred(i)= \{j \in V_r | (j,i) \in {\mathcal A}_r \}$.

We say that a node is a sink if it has no successors on the graph. For the sake of simplicity,
we suppose that there is only one sink. For each one of the remaining $n-1$ nodes there is
one reaction leaving from it. For a network with totally separated constants we have
\begin{equation}
\kappa_i \ll   \kappa_j, \text{ or }  \kappa_i \gg   \kappa_j \text{ for all } i,j \in [1,n-1],\, i\neq j
\label{separation}
\end{equation}
For totally separated constants the following lemma is useful
\begin{lemma}
\label{totalseplemma}
If \eqref{separation} is satisfied then, at lowest order, we have
\begin{equation}
\frac{\kappa_i}{-\kappa_k + \kappa_j} =
\left\{
\begin{array}{ll}
1, & \text{if } i=j \text{ and } \kappa_k < \kappa_i \\
-1, & \text{if } i=k \text{ and } \kappa_j < \kappa_i \\
0, & \text{if } \kappa_i < \min (\kappa_k,\kappa_j) \\
\pm \infty, & \text{else}
\end{array}
\right.
\end{equation}
\end{lemma}
The dynamics of the reduced model is given by
\begin{equation}\label{kinurred}
\D{c_i}{ t}=\sum_{j \in \Pred(i)} \kappa_{j} c_j - \kappa_i c_i,
\end{equation}
where $\Pred(i)$ is the set of predecessors of the node $i$ in the digraph $G_r$,
namely $\Pred(i)= \{j \in V_r | (j,i) \in {\mathcal A}_r \}$.

As shown in \cite{gorban-dynamic} the eigenvectors of the approximated kinetic matrix satisfy
\begin{eqnarray}
\sum_{j \in \Pred(i)} \kappa_{j} r_j &=& (\lambda + \kappa_i) r_i  \label{recurrence1}\\
 \kappa_{i} l_{\Phi(i)} &=& (\lambda + \kappa_i) l_i,
 \label{recurrence2}
\end{eqnarray}
where $\lambda$ is the eigenvalue, $r_i$, $l_i$, $1\leq i \leq n$ are the components of the right and left eigenvectors, respectively.

Eqs.(\ref{recurrence1}) and (\ref{recurrence2}) imply that the right and left eigenvectors can be computed
by recurrence on the graph, in the direct direction and in the reverse direction,
respectively.
In order to have non-zero eigenvectors, $\lambda = - \kappa_i$ for some $i$ not a sink,
therefore the (non-zero) eigenvalues are $\lambda_k  = - \kappa_k, 1 \leq k \leq n-1$.
Taking into account the separation conditions \eqref{separation} we get the following
\begin{proposition} \label{eigenvectors}
Let us consider that $\kappa_k=0$ when $k$ is a sink in the graph $G_r$.
Then, the eigenvalues of the kinetic matrix with totally separated constants
are $\lambda_k = - \kappa_k$, with $\lambda_k=0$ when $k$ is a sink.
The corresponding left eigenvectors are
\begin{equation}
l_j^k =
\left\{
\begin{array}{ll}
1, & \text{if } \Phi^m(j)=k \text{ for some } m>0 \text{ and }
\kappa_{\Phi^l(j)}>\kappa_k \text{ for all } l=0,\ldots,m-1 \\
0, & \text{otherwise}
\end{array}
\right.,
\end{equation}
and the right eigenvectors are
\begin{equation}
r_j^k =
\left\{
\begin{array}{ll}
1, & \text{if } j=k \\
-1, &
\text{ if }
j = \Phi^m(k) \text{ for some } m>0
\text{ and }
\kappa_{\Phi^{m}(k)} < \kappa_k
< \kappa_{\Phi^l(k)}, \notag \\
& \text{ for all }
l = 1,\ldots,m-1
 \\
0, & \text{otherwise.}
\end{array}
\right.
\end{equation}
\end{proposition}
The full proof of the Proposition~\ref{eigenvectors} can be found in the appendices.

Let us now discuss the symbolic dynamics of the system.
For each eigenvalue $\lambda_k = - \kappa_k$, $\kappa_k > 0$ we associate a transition
time $t_k = \kappa_k^{-1}$. Without loss of generality we can consider
that
$t_1 \ll  t_2 \ll  \ldots \ll  t_{n-1}.$
Any trajectory of the system is given by \eqref{solkinur}.
At the time $t_k$ one exponential term $exp(\lambda_k t)$ will
vanish and the result will be a transition $c \rightarrow c - \vect{r}^k (\vect{l}^k, \vect{c}(0))$,
provided that  $(\vect{l}^k, \vect{c}(0)) \neq 0$.
In other works, a trajectory can be described as a discrete
sequence of states $c(0), c(0) - \vect{r}^1 (\vect{l}^1, \vect{c}(0)), \ldots$.
Let us consider the following normalization
$c_1(0) + c_2(0) + \ldots + c_n(0) =1$. Then $c_i$ is the probability
of presence in the node $i$ of a particle moving through the reaction network. For
monomolecular networks, particles are independent, therefore this simple picture is
enough for understanding the dynamics.
Let the index $i_0$ define the initial state of the system
$c_{i_0}(0)=1$, $c_{j}(0)=0$ for $j\neq i_0$. $i_0$ represents
the initial position of the particle.
According to the Prop. \ref{eigenvectors}
$(\vect{l}^k, \vect{c}(0)) = l^k_{i0} =1$ if the step $\kappa_k$
is downstream of $i_0$ in the graph $G_r$ and if all steps from $i_0$ to $k$
are faster than $\kappa_k$. In this case the jump at $t_k$ is $-\vect{r}^k$.
A jump $-\vect{r}^k$ has two components different from zero,
$-r^k_k=-1$ and $-r^k_j=1$, where $j$ is the first node downstream of $k$ from which
starts a step slower than $\kappa_k$. Thus, the jump $-\vect{r}^k$ corresponds to displacing the
particle from $k$ to $j$.  The set of right eigenvectors defines
a symbolic flow on the reaction digraph.
A particle starting in $i_0$
first jumps in $i_1$ where $i_1$ is the first node
such that $\kappa_{i_1} < \kappa_{i_0}$, then continues to $i_2$
where $i_2$ is the first node such that
$\kappa_{i_2} < \kappa_{i_1}$, and
so one and so forth until it gets to the sink.
Some nodes have negligible sojourn time, namely nodes such that
$\kappa_i > \kappa_j \text{ for all } j \in \Pred(i)$.
This proves the main result of the section.

By transition graph of a finite state machine we mean the digraph
$G_{rs} = (V_s, {\cal A}_s)$, where $V_s$ is the set of states of the machine and
$(i,j) \in {\cal A}_s$ if there are transitions from the state $i$ to the state $j$.
We have the following theorem:
\begin{theorem}
The symbolic dynamics of a monomolecular network with totally separated constants can be
described by a deterministic acyclic finite state machine. The  transition graph
$G_{rs} = (V_s, {\cal A}_s)$
of this
machine can be obtained from the graph $G_r = (V_r, {\cal A}_r)$ in the following way:
$V_s = V_r \setminus \{i \in V_r | \kappa_i > \kappa_j \text{ for all } j \in \Pred(i) \}$,
${\cal A}_s = \{ (i,j) | i,j \in V_s \text{ and there are } i_0=i,i_1,\ldots,i_m=j, \text{ such that }
i_l \in  V_r\setminus V_s, \text{ for } l=1,\ldots,m-1, \text{ and }
(i_l, i_{l+1})\in {\cal A}_r \text{ for } l=0,\ldots,m-1   \}$.
\end{theorem}
\begin{remark}
An example is detailed in Figure~\ref{fig1}.
\end{remark}
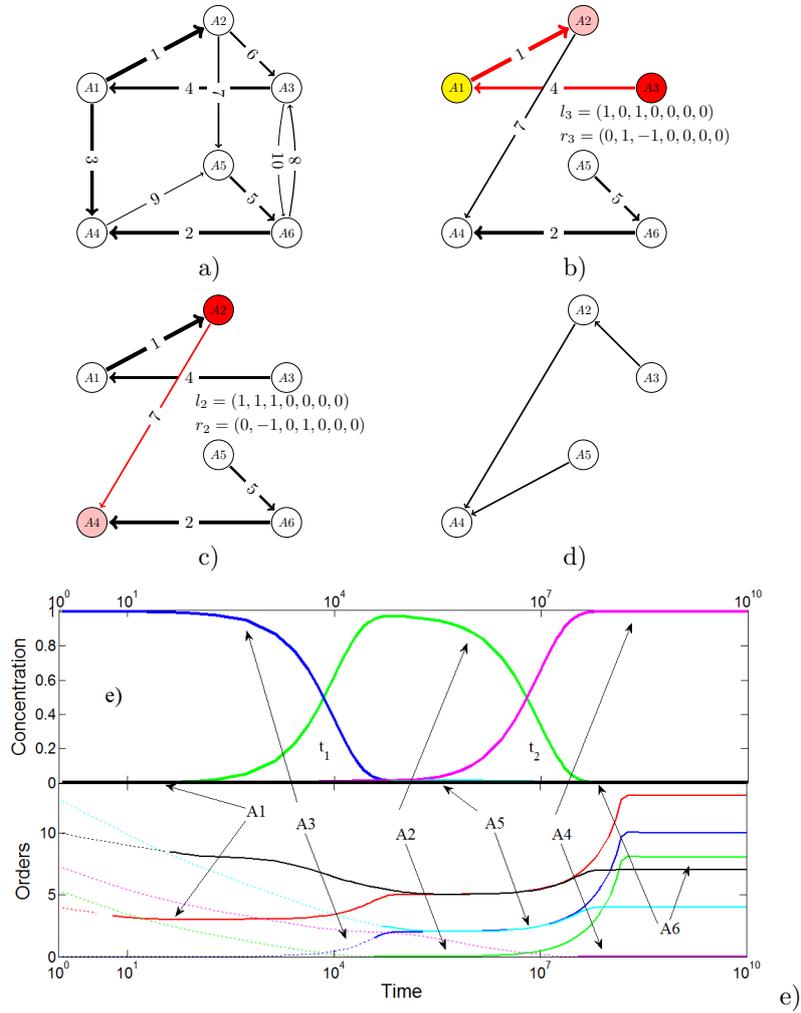
\begin{figure}
\begin{center}
\scalebox{0.32}{
\begin{tikzpicture}
 \SetUpEdge[lw         = 0.5pt,
            color      = black,
            labelstyle = {sloped,scale=2}]
  \SetVertexMath
       \tikzset{VertexStyle/.style={scale=1.5,
       draw,
            shape = circle,
            line width = 1pt,
            color = black,
            outer sep=1pt}}
  \Vertex[x=0,y=6]{A1}
  \Vertex[x=4+1.2,y=6+2.8]{A2}
  \Vertex[x=8,y=6]{A3}
  \Vertex[x=0,y=0]{A4}
  \Vertex[x=4+1.2,y=2.8]{A5}
  \Vertex[x=8,y=0]{A6}

\tikzset{EdgeStyle/.style={->, line width = 5}}
\Edge[label=$1$](A1)(A2)

\tikzset{EdgeStyle/.style={post,line width = 2.5}}
\Edge[label=$6$](A2)(A3)

\tikzset{EdgeStyle/.style={post,line width = 3.5}}
\Edge[label=$4$](A3)(A1)

\tikzset{EdgeStyle/.style={post,line width = 1}}
\Edge[label=$9$](A4)(A5)

\tikzset{EdgeStyle/.style={post,line width = 3}}
\Edge[label=$5$](A5)(A6)

\tikzset{EdgeStyle/.style={post,line width = 4.5}}
\Edge[label=$2$](A6)(A4)

\tikzset{EdgeStyle/.style={post,line width = 4}}
\Edge[label=$3$](A1)(A4)

\tikzset{EdgeStyle/.style={post,line width = 2}}
\Edge[label=$7$](A2)(A5)

\tikzset{EdgeStyle/.style={post, thick, bend right = 10,line width = 0.5}}
\Edge[label=$10$](A3)(A6)

\tikzset{EdgeStyle/.style={post,bend right = 10,line width = 1.5}}
\Edge[label=$8$](A6)(A3)

\draw(4,-1.5)node[above,right, scale =3] {a)};

\begin{scope}[xshift=15cm]

 \SetVertexMath
       \tikzset{VertexStyle/.style={scale=1.5,
       draw,
            shape = circle,
            line width = 1pt,
            color = black,
            fill = yellow,
            outer sep=1pt}}

\Vertex[x=0,y=6]{A1}

     \tikzset{VertexStyle/.style={scale=1.5,draw,
            shape = circle,
            line width = 1pt,%
            color = black,%
            fill = pink,outer sep=1pt}}

  \Vertex[x=4+1.2,y=6+2.8]{A2}

\tikzset{VertexStyle/.style={scale=1.5,draw,
            shape = circle,
            line width = 1pt,%
            color = black,%
            fill = red,outer sep=1pt}}
  \Vertex[x=8,y=6]{A3}

 \SetVertexMath
       \tikzset{VertexStyle/.style={scale=1.5,
       draw,
            shape = circle,
            line width = 1pt,
            color = black,
            outer sep=1pt}}

  \Vertex[x=0,y=0]{A4}
  \Vertex[x=4+1.2,y=2.8]{A5}
  \Vertex[x=8,y=0]{A6}

\tikzset{EdgeStyle/.style={->, line width = 5, color = red}}
\Edge[label=$1$](A1)(A2)

\tikzset{EdgeStyle/.style={post,line width = 3.5, color = red}}
\Edge[label=$4$](A3)(A1)

\tikzset{EdgeStyle/.style={post,line width = 3}}
\Edge[label=$5$](A5)(A6)

\tikzset{EdgeStyle/.style={post,line width = 4.5}}
\Edge[label=$2$](A6)(A4)

\tikzset{EdgeStyle/.style={post,line width = 2}}
\Edge[label=$7$](A2)(A4)

\draw(4,5)node[above,right,scale=2] {$l_3=(1,0,1,0,0,0,0)$ };
\draw(4,4)node[above,right, scale=2] {$r_3=(0,1,-1,0,0,0,0)$};
\draw(4,-1.5)node[above,right, scale =3] {b)};
 \end{scope}

\begin{scope}[yshift=-12cm]

 \SetVertexMath
       \tikzset{VertexStyle/.style={scale=1.5,
       draw,
            shape = circle,
            line width = 1pt,
            color = black,
            outer sep=1pt}}

\Vertex[x=0,y=6]{A1}
\Vertex[x=8,y=6]{A3}
  \Vertex[x=4+1.2,y=2.8]{A5}
  \Vertex[x=8,y=0]{A6}

\tikzset{VertexStyle/.style={scale=1.5,draw,
            shape = circle,
            line width = 1pt,%
            color = black,%
            fill = red,outer sep=1pt}}
\Vertex[x=4+1.2,y=6+2.8]{A2}
\tikzset{VertexStyle/.style={scale=1.5,draw,
            shape = circle,
            line width = 1pt,%
            color = black,%
            fill = pink,outer sep=1pt}}
  \Vertex[x=0,y=0]{A4}

\tikzset{EdgeStyle/.style={->, line width = 5}}
\Edge[label=$1$](A1)(A2)

\tikzset{EdgeStyle/.style={post,line width = 3.5}}
\Edge[label=$4$](A3)(A1)

\tikzset{EdgeStyle/.style={post,line width = 3}}
\Edge[label=$5$](A5)(A6)

\tikzset{EdgeStyle/.style={post,line width = 4.5}}
\Edge[label=$2$](A6)(A4)

\tikzset{EdgeStyle/.style={post,line width = 2, color = red}}
\Edge[label=$7$](A2)(A4)

\draw(4,5)node[above,right,scale=2] {$l_2=(1,1,1,0,0,0,0)$ };
\draw(4,4)node[above,right, scale=2] {$r_2=(0,-1,0,1,0,0,0)$};
\draw(4,-1.5)node[above,right, scale =3] {c)};
 \end{scope}

\begin{scope}[xshift=15cm,yshift=-12cm]

 \SetVertexMath
       \tikzset{VertexStyle/.style={scale=1.5,
       draw,
            shape = circle,
            line width = 1pt,
            color = black,
            outer sep=1pt}}

\Vertex[x=4+1.2,y=6+2.8]{A2}
\Vertex[x=8,y=6]{A3}
\Vertex[x=0,y=0]{A4}
\Vertex[x=4+1.2,y=2.8]{A5}

\tikzset{EdgeStyle/.style={->, line width = 2}}
\Edge(A3)(A2)
\Edge(A2)(A4)
\Edge(A5)(A4)

\draw(4,-1.5)node[above,right, scale =3] {d)};
 \end{scope}

\end{tikzpicture}
}

\scalebox{0.50}{\includegraphics[width=20cm]{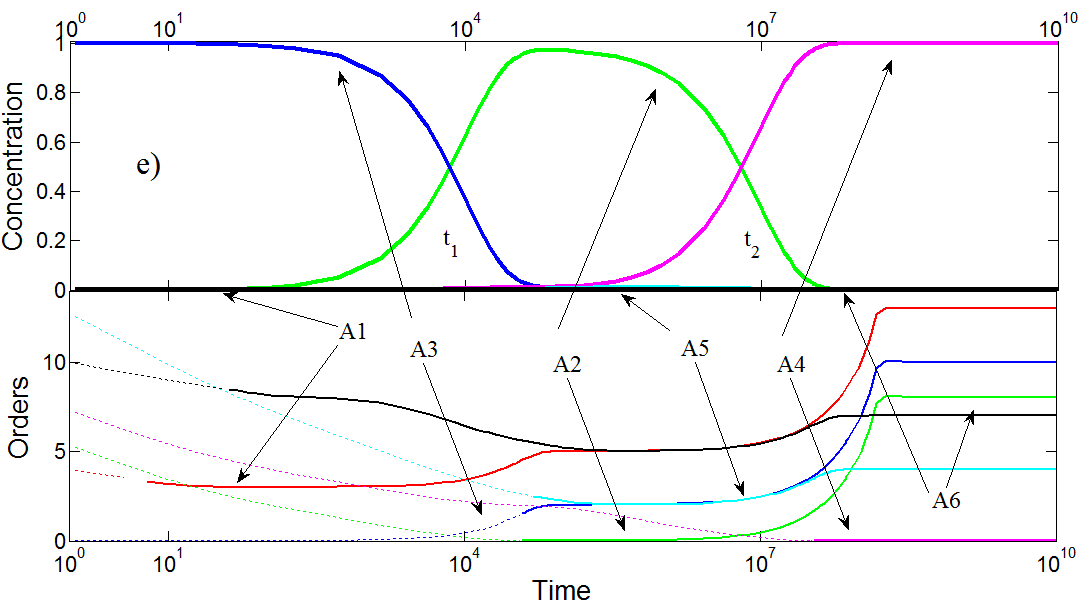}}
\vspace*{-2mm} e)
\end{center}
\caption{Symbolic dynamics of a monomolecular network with total separation.
The integers $\gamma_i$ labelling the
reactions represent the orders of the kinetic constants, smaller orders
meaning faster reactions. The model was reduced using the recipe described in
\cite{gorban-dynamic,radulescu2012frontiers} (see appendices).
a) full model; b-c) reduced model with active transitions and
corresponding eigenvectors. During a transition the network
behaves like a single step : the concentrations of some species (white)
are practically constant, some species (yellow) are rapid,
low concentration, intermediates, one species (red) is gradually
consumed and another (pink) is gradually produced.
The net result is the displacement of a particle one or several steps downstream;
d) The transition graph of the finite state machine representing the symbolic dynamics of the
network;
e) Trajectory starting from $A3$ (at $t=0$ the total mass is in $A3$), undergoing two transitions at $t_1$ and $t_2$. The simulation has been performed
for kinetic constants $\kappa_i = \varepsilon^{\gamma_i}$, with $\varepsilon = 1/50$.
On top, concentration of species (concentrations of $A1$,$A_4$,$A_6$ are negligible everywhere). At bottom, orders of concentrations (computed as $\log_{\epsilon}(x_i)$) with
continuous lines if species is tropically equilibrated, dotted lines if not. }
\label{fig1}
\end{figure}

\section{Tropical equilibrations of nonlinear networks with polynomial rate functions}
In this section we consider nonlinear biochemical networks described by mass action kinetics
 \begin{equation}
 \D{x_i}{t} = \sum_j k_j S_{ij}  \vect{x}^{\vect{\alpha_{j}}}, \, 1 \leq i \leq n,
 \label{massaction}
 \end{equation}
where $k_j >0$ are kinetic constants, $S_{ij}$ are the
entries of the stoichiometric matrix (uniformly bounded integers,
$|S_{ij}| < s$, $s$ is small),
$\vect{\alpha}_{j} = (\alpha_1^j, \ldots, \alpha_n^j)$ are multi-indices,
  and $\vect{x^{\vect{\alpha}_{j}}}  = x_1^{\alpha_1^j} \ldots x_n^{\alpha_n^j}$, where $\alpha_i^j$ are positive integers.

For chemical reaction networks with multiple timescales it is reasonable to consider
that kinetic parameters have different orders of magnitudes. This can be conveniently
formalized by considering
 that parameters of the kinetic models
\eqref{massaction} can be written as
\inlineequation[scaleparam]{
k_j = \bar k_j \varepsilon^{\gamma_j}}.  
The exponents $\gamma_j$ are considered to be integer or rational. For instance, the
approximation
$\gamma_j = \text{round}( \log(k_j) / \log(\varepsilon))$
produces integer exponents,
 whereas
$\gamma_j =  \text{round}( d \log(k_j) / \log(\varepsilon)) / d$ produces rational exponents,
where round stands for the closest integer (with half-integers rounded to even numbers) and
$d$ is a strictly positive integer.
Kinetic parameters are fixed. In contrast, species orders vary
 in the concentration space and have to be calculated as solutions to the tropical equilibration problem.
To this aim, the network dynamics is first described by a rescaled ODE system
 \begin{equation}
 \D{\bar{x}_i}{t} = \sum_j \varepsilon^{\mu_j(\vect{a}) - a_i} \bar k_j S_{ij}   {\bar{\vect{x}}}^{\vect{\alpha_{j}}},
 \label{massactionrescaled}
 \end{equation}
where
\inlineequation[muj]{
\mu_j(\vect{a}) = \gamma_j +  \langle \vect{a},\vect{\alpha_j}\rangle },
and $\langle , \rangle $ stands for the dot product. 

The r.h.s.\ of each equation in
\eqref{massactionrescaled} is a sum of multivariate monomials in the concentrations.
The orders $\mu_j$ indicate how large are these monomials, in absolute value.
A monomial of order $\mu_j$ dominates another monomial of order
$\mu_{j'}$ if
 $\mu_j < \mu_{j'}$.

{\em The tropical equilibration problem} consists in the equality of the orders of
at least two monomials one positive and another negative in the differential
equations of each species. More precisely, we want to find a vector $\vect{a}$ such that
\begin{equation}
\min_{j,S_{ij}  >0} ( \gamma_j + \langle \vect{a},\vect{\alpha_j}\rangle ) =
\min_{j,S_{ij}  <0} ( \gamma_j + \langle \vect{a},\vect{\alpha_j}\rangle )
\label{eq:minplus}
\end{equation}
Computing tropical equilibrations from the orders of magnitude of the model parameters
is a NP-complete problem, cf. \cite{theobald2006frontiers}. However,
methods based on the Newton polytope \cite{samal2014tropical} or constraint logic programming \cite{soliman2014constraint} exploit the sparseness and redundance of
the system to effectively obtain sets of solutions.
The equation\eqref{eq:minplus} is related to the notion
of {\em tropical hypersurface}. A {\em tropical hypersurface} is the set of vectors
$\vect{a} \in \R^n$ such that the minimun $\min_{j,S_{ij}  \neq 0} ( \gamma_j + \langle \vect{a},\vect{\alpha_j}\rangle )$ is attained for at least two different indices $j$
(with no sign conditions). {\em Tropical prevarieties} are finite intersections of
tropical hypersurfaces. Therefore, our tropical equilibrations are subsets of
tropical preverieties. The sign condition in  \eqref{eq:minplus} was imposed
because species concentrations are real positive numbers. Compensation
of a sum of positive monomials is not possible for real values of the variables.

{\em Species timescales.}
The timescale of a variable $x_i$ is given by $ \frac{1}{x_i}\D{x_i}{t} = \frac{1}{\bar{x}_i}\D{\bar{x}_i}{t}$
whose order is
\inlineequation[timescales]{\nu_{i}  = \min \{ \mu_j |  S_{ij} \neq 0 \} - a_i}.
The order  $\nu_{i}$ indicates how fast is the variable $x_i$ (if
$\nu_{i'} < \nu_{i}$ then $x_{i'}$ is faster than $x_{i}$) .

{\em Partial tropical equilibrations.}
It is useful to extend the tropical equilibration problem to partial equilibrations,
that means solving \eqref{eq:minplus} only for a subset of species. This is justified
by the fact that slow species do not need to be equilibrated. In order to have a
self-consistent calculation we compute the species timescales by \eqref{timescales}.
A partial equilibration is {\em consistent} if $\nu_i < \nu$ for all non-equilibrated
species $i$. $\nu > 0$ is an arbitrarily chosen threshold indicating the timescale of
interest.

{\em Tropical equilibrations, slow invariant manifolds and metastable states.}
In dissipative systems, fast variables relax  rapidly to some low dimensional attractive
 manifold called invariant manifold \cite{gorban2005invariant} that carries the slow mode  dynamics.
 A projection of dynamical equations
 onto this manifold provides the reduced dynamics \cite{maas1992simplifying}.
 This simple picture can be complexified to cope with hierarchies of invariant manifolds and
 with phenomena such as transverse instability, excitability and itineracy.
Firstly, the relaxation towards an attractor can have several stages, each with its own invariant manifold.
 During relaxation towards the attractor, invariant manifolds are usually embedded one into another (there is a decrease of dimensionality) \cite{chiavazzo2011adaptive}.
Secondly, invariant manifolds can lose local stability, which allow the trajectories to perform
large phase space excursions before returning in a different place on the same invariant manifold or on
a different one \cite{haller2010localized}.
We showed elsewhere that tropical equilibrations can be used to approximate
invariant manifolds for systems of polynomial differential
equations \cite{NGVR12sasb,Noel2013a,Radulescu2015}.
Indeed,
tropical equilibration are defined by the equality of dominant forces acting on the system. The remaining
weak non-compensated forces ensure the slow dynamics on the invariant manifold.
Tropical equilibrations are thus different from steady states, in that there is a slow dynamics.
In this paper we will use them as proxies for metastable states.

{\em Branches of tropical equilibrations and connectivity graph.}
For each equation $i$, let us define
\inlineequation[Mi]{M_i(\vect{a}) = \underset{j}{\argmin}
 (\mu_j(\vect{a}), S_{ij}  >0) = \underset{j}{\argmin} (\mu_j(\vect{a}),S_{ij}  <0)},
in other words
$M_i$ denotes the set of monomials having the same minimal order $\mu_i$.
We call {\em tropically truncated system} the system obtained by pruning the system
\eqref{massactionrescaled}, i.e. by keeping only the dominating monomials.
 \begin{equation}
 \D{\bar{x}_i}{t} = \varepsilon^{\mu_i - a_i} (\sum_{j\in M_i (\vect{a})} \bar k_j  \nu_{ji}  {\bar{\vect{x}}}^{\vect{\alpha_{j}}}),
 \label{massactionrescaledtruncated}
 \end{equation}
The tropical truncated system is uniquely determined by the index sets
$M_i(\vect{a})$, therefore by the
tropical equilibration
$\vect{a}$. Reciprocally, two tropical equilibrations can have the same index sets
$M_i(\vect{a})$ and truncated systems. We say that two tropical equilibrations $\vect{a}_1$, $\vect{a}_2$
are  equivalent iff $M_i(\vect{a}_1) = M_i(\vect{a}_2), \text{for all } i$. Equivalence classes
of tropical equilibrations
are called
{\em branches}. A branch $B$ with an index set $M_i$  is {\em minimal} if
$M'_i \subset M_i$ for all $i$ where $M'_i$ is the index set $B'$ implies
$B'=B$ or $B'=\emptyset$.
Closures of equilibration branches are defined by a finite set of
linear inequalities, which means that they are polyhedral complexes.
Minimal branches correspond to maximal
dimension faces of the polyhedral complex.
The incidence relations between the maximal dimension faces ($n-1$ dimensional faces,
where $n$ is the number of variables) of the polyhedral complex define
the {\em connectivity graph}. More precisely, minimal branches are the vertices of this graph. Two minimal branches are connected if the corresponding faces
of the polyhedral complex share a $n-2$ dimensional face. In terms of index sets,
two minimal branches with index sets $M$  and $M'$ are connected if there is
an index set $M''$ such that $M'_i \subset M''_i$ and $M_i \subset M''_i$ for all $i$.

{\em Tropical equilibrations and monomolecular networks.}
Eqs.\eqref{eq:minplus} have a simpler form in the case of monomolecular networks
\begin{equation}
\min_{j \in \Pred(i)} ( \gamma_{ij} + a_j ) =
\min_{j \in \Succ(i)} ( \gamma_{ji} + a_i )
\label{eq:linminplus}
\end{equation}
where $\Pred(i) = \{j | (j,i) \in {\mathcal A} \}$, $\Succ(i) = \{j | (i,j) \in {\mathcal A} \}$ are
the sets of predecessors and successors of the node $i$ in the digraph $G$.

Let us recall that by min-plus algebra we understand the semi-ring $(\R \cup \{ \infty \},\oplus,\otimes)$
where the two operations are defined as $x \oplus y = \min\{ x,y \}$ and
 $x \otimes y = x + y$. In other words the addition and
 the $\min$ operation play the role of min-plus multiplication and
addition, respectively. Therefore Eqs.\eqref{eq:linminplus} are
linear in the unknowns $a_i$. Computing tropical equilibrations of
monomolecular networks boils down to solving linear equations
in min-plus algebra. For linear tropical systems there are fast
algorithms \cite{grigoriev2013complexity,grigoriev2014complexity}.

We have tested the tropical equilibration conditions
\eqref{eq:linminplus} for the trajectories of the
monomolecular network presented in Figure~\ref{fig1}
by checking if the absolute value of the difference between the r.h.s and
l.h.s of \eqref{eq:linminplus} is smaller than a threshold. The result
is illustrated in Fig.~\ref{fig1}e). For this model, the tropical
equilibration solutions are changing along the trajectory. This can
been seen by following the orders of the concentrations along the
trajectories. These orders change by integers at transition points.
Furthermore,
at transition points some of the variables that where not
previously equilibrated, become equilibrated.
The analysis of the tropical equilibrations finds the transitions
previously detected in Section~\ref{sec:mono} from the approximated eigenvalues and eigenvectors
($t_1$ and $t_2$ for this example) but adds some more. For instance, species
$A1$ equilibrates at the timescale $1/\kappa_1 = 10$. This was not taken into
account in the description of the automaton in Figure~1d) because the species $A1$ is fast
and can not accumulate.
\section{Learning a finite state machine from a nonlinear biochemical network}
We are using the algorithm based on constraint solving
introduced in \cite{soliman2014constraint} to
obtain all rational tropical equilibration solutions $\vect{a} = (a_1,a_2,\ldots,a_n)$
within a box $|a_i| < b$, $b >0$ and with denominators smaller than a fixed value
$d$, $a_i = p_i/q$, $p_i,q$ are positive integers, $q < d$.
The output of the algorithm is a matrix containing all the tropical equilibrations within
the defined bounds. A post-processing treatment is applied to this output
consisting in computing truncated systems, index sets, and minimal branches.
Tropical equilibrations minimal branches are stored as matrices $A_1,A_2,\ldots,A_b$, whose lines are
tropical solutions within the same branch. Here $b$ is the number of minimal branches.

Our method computes numerical approximations of the tropical prevariety. Given a value of
$\epsilon$, this approximation is better when the denominator bound $d$ is high. At fixed
$d$, the dependence of the precision on $\epsilon$ follows more intricate rules dictated by
Diophantine approximations. For this reason, we systematically test
that the number $b$ and the truncated systems corresponding
to minimal branches are robust when changing the value of $\epsilon$.

Trajectories $\vect{x}(t) = (x_1(t),\ldots,x_n(t))$ of the smooth dynamical system are
generated with different initial conditions, chosen uniformly and satisfying the conservation laws, if any.
For each time $t$, we compute the Euclidian distance
$d_i(t) = \min_{\vect{y} \in A_i}  \left \lVert \vect{y} - log_{\varepsilon}(\vect{x}(t)) \right \rVert,$
where
$\left \lVert * \right \rVert$ denotes the Euclidean norm and
$\log_\varepsilon(\vect{x}) = (\log x_1 / \log(\varepsilon) ,\ldots, \log x_n / \log(\varepsilon))$.
This distance classifies all points of the
trajectory as belonging to a tropical minimal branch. The result is a symbolic trajectory
$s_1,s_2,\ldots$ where the symbols $s_i$ belong to the set of minimal branches.
In order to include the possibility of transition regions we include an unique symbol $t$
to represent the situations when the minimal distance is larger than a fixed threshold.
We also store the residence times $\tau_1,\tau_2,\ldots$ that represent the
time spent in each of the state.

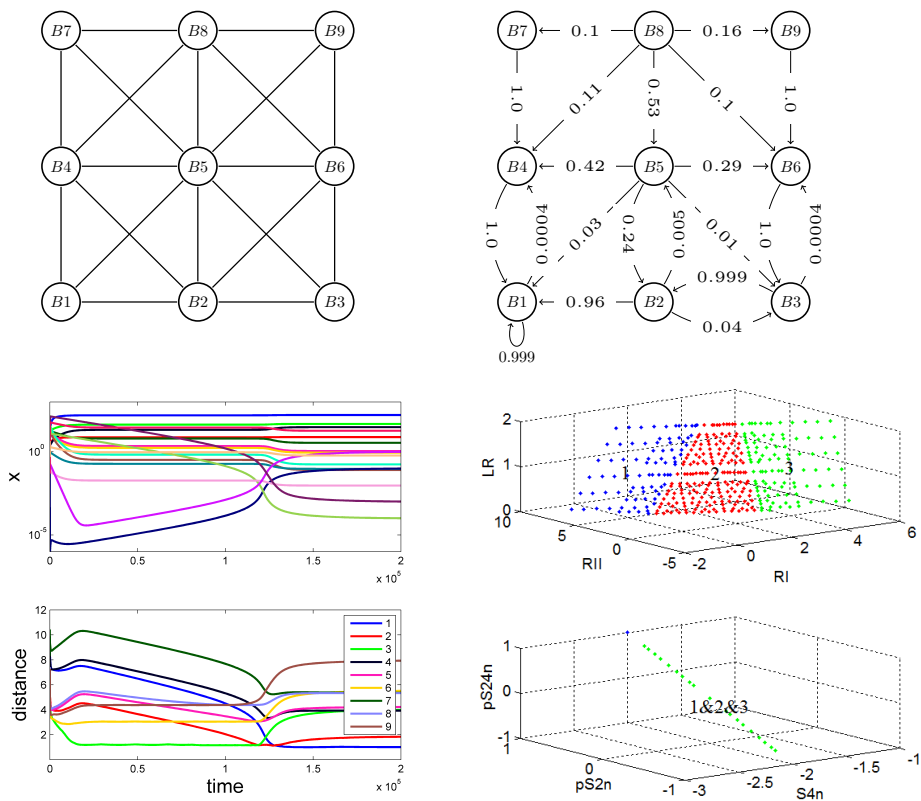
\begin{figure}[h!]
\begin{center}
\scalebox{0.6}{
\vspace*{-5mm}
\begin{tikzpicture}
 \SetUpEdge[lw         = 0.5pt,
            color      = black,
            labelstyle = {sloped,scale=2}]
  \tikzset{node distance = 1.5cm}
  \GraphInit[vstyle=Normal]
  \SetVertexMath
       \tikzset{VertexStyle/.style={scale=1.0,
       draw,
            shape = circle,
            line width = 1pt,
            color = black,
            outer sep=1pt}}
  \Vertex[x=0,y=0]{B1}
  \Vertex[x=3,y=0]{B2}
  \Vertex[x=6,y=0]{B3}
  \Vertex[x=0,y=3]{B4}
  \Vertex[x=3,y=3]{B5}
  \Vertex[x=6,y=3]{B6}
    \Vertex[x=0,y=6]{B7}
  \Vertex[x=3,y=6]{B8}
  \Vertex[x=6,y=6]{B9}

  \Edge(B1)(B2)
  \Edge(B2)(B3)
    \Edge(B4)(B5)
  \Edge(B5)(B6)
    \Edge(B7)(B8)
  \Edge(B8)(B9)
    \Edge(B1)(B4)
  \Edge(B2)(B5)
    \Edge(B3)(B6)
  \Edge(B4)(B7)
    \Edge(B5)(B8)
  \Edge(B6)(B9)
    \Edge(B1)(B5)
  \Edge(B2)(B4)
  \Edge(B3)(B5)
  \Edge(B2)(B6)
    \Edge(B4)(B8)
  \Edge(B5)(B7)
    \Edge(B6)(B8)
  \Edge(B5)(B9)

\begin{scope}[xshift=10cm]
\SetUpEdge[lw         = 0.5pt,
            color      = black,
            labelstyle = {sloped,scale=2}]
  \tikzset{node distance = 1.5cm}
  \GraphInit[vstyle=Normal]
  \SetVertexMath
       \tikzset{VertexStyle/.style={scale=1.0,
       draw,
            shape = circle,
            line width = 1pt,
            color = black,
            outer sep=1pt}}
  \Vertex[x=0,y=0]{B1}
  \Vertex[x=3,y=0]{B2}
  \Vertex[x=6,y=0]{B3}
    \Vertex[x=0,y=3]{B4}
  \Vertex[x=3,y=3]{B5}
  \Vertex[x=6,y=3]{B6}
    \Vertex[x=0,y=6]{B7}
  \Vertex[x=3,y=6]{B8}
  \Vertex[x=6,y=6]{B9}
\tikzset{EdgeStyle/.style={post,line width = 0.5, font=\tiny}}
\Edge[label=$0.96$](B2)(B1)
\Edge[label=$0.03$](B5)(B1)
\Edge[label=$0.42$](B5)(B4)
\Edge[label=$0.01$](B5)(B3)
\Edge[label=$0.29$](B5)(B6)
\Edge[label=$0.16$](B8)(B9)
\Edge[label=$0.11$](B8)(B4)
\Edge[label=$0.1$](B8)(B6)
\Edge[label=$0.53$](B8)(B5)
\Edge[label=$0.1$](B8)(B7)
\Edge[label=$1.0$](B7)(B4)
\Edge[label=$1.0$](B9)(B6)
\path (B1) edge [loop below] node {0.999} (B1);

\tikzset{EdgeStyle/.style={post,bend right,line width = 0.5, font=\tiny}}
\Edge[label=$0.0004$](B1)(B4)
\Edge[label=$1.0$](B4)(B1)
\Edge[label=$0.005$](B2)(B5)
\Edge[label=$0.24$](B5)(B2)
\Edge[label=$0.0004$](B3)(B6)
\Edge[label=$1.0$](B6)(B3)
\Edge[label=$0.04$](B2)(B3)
\Edge[label=$0.999$](B3)(B2)

\end{scope}
  \end{tikzpicture}
}
\vspace*{-5mm}
\begin{tabular}[b]{cc}
 \scalebox{0.33}{\includegraphics[width=18cm]{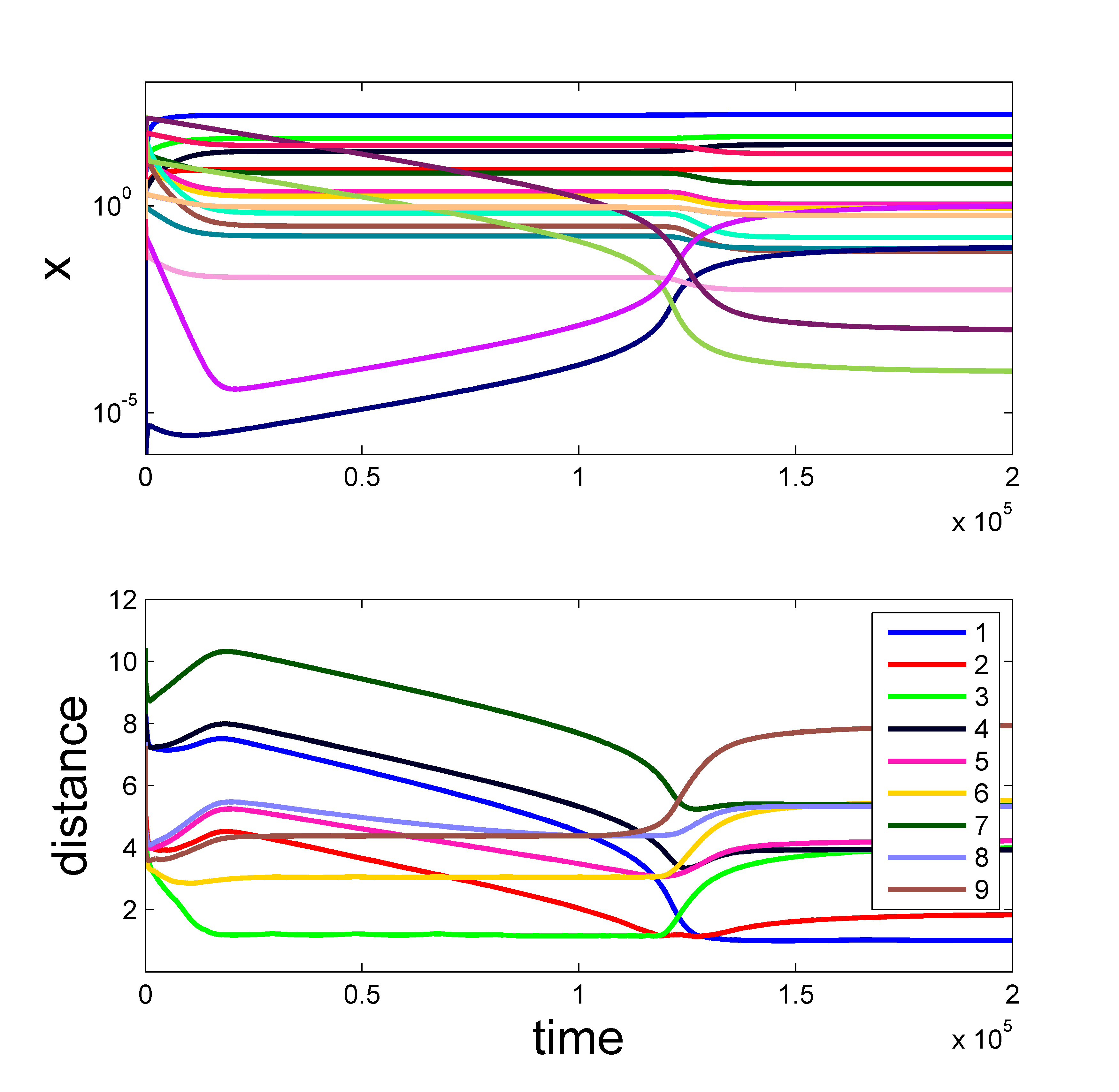}} &  \scalebox{0.33}{\includegraphics[width=18cm]{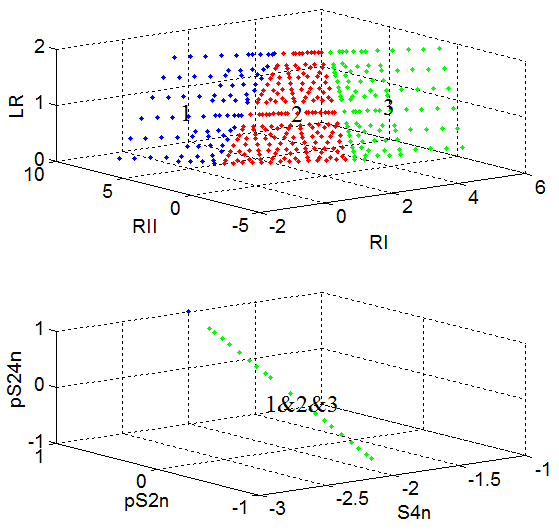}}
\end{tabular}
\end{center}
\caption{\label{example2}
TGF$\beta$ model. Upper left: Connectivity graph of tropical minimal branches; upper right: finite
state automaton; bottom left: trajectories with jumps and distances to minimal branches; the closest
branch changes with time along the trajectory; bottom right: first three tropical equilibrations minimal
branches in various projections in concentration orders space. The variables RI, RII, LR are membrane receptors
(signaling input layers) concentration orders, whereas pS2n, S4n, pS24n are nuclear transcription factors
and complexes (effectors) concentration orders. The structure tropical branches shows that composition of input layers is more flexible (varies on planes) than the concentrations of effectors (vary on lines). }
\end{figure}

The stochastic automaton is learned as a homogenous, finite states, continuous time Markov process,
defined by the lifetime (mean sojourn time) of each state $T_i$, $1 \leq i\leq b$ and
by the transition probabilities $p_{i,j}$ from a state $i$ to another state $j$.
We use the following estimators for the lifetimes and for the transition probabilities:
\begin{eqnarray}
T_i  &=& (\sum_{n} \tau_n  \one_{s_n = i}) / (\sum_{n} \one_{s_n = i}) \\
p_{i,j} &=& (\sum_{n} \one_{ s_n = i, s_{n+1} = j} ) / (\sum_{n}  \one_{s_n = i}), \, i \neq j
\end{eqnarray}
As a case study we consider a nonlinear model of dynamic regulation of Transforming Growth Factor beta TGF-$\beta$ signaling pathway proposed in \cite{andrieux2012dynamic}. This model has a dynamics defined by $n=18$ polynomial differential equations and  $25$ biochemical reactions.
The paper  \cite{andrieux2012dynamic} proposes three versions of the
mechanism of interaction of TIF1$\gamma$ (Transcriptional Intermediary Factor 1 $\gamma$) with the  Smad-dependent TGF-$\beta$ signaling. We consider here the
version in which TIF1 interacts with the phosphorylated  Smad2--Smad4  complexes
leading to dissociation of the complex and degradation of Smad4. The results are similar for the
other versions of this model. The example was chosen because it is a medium size model based on
polynomial differential equations. The computation of the tropical equilibrations for this model shows that there are $9$ minimal
branches of full equilibrations (in these tropical solutions all variables are equilibrated).
The connectivity graph of these branches  and the learned
automaton are shown in Figure~\ref{example2}. The study of this example shows that branches of tropical equilibration can change on trajectories of the dynamical system. Furthermore,
all the observed transitions between branches are contained in the connectivity graph
resulting from the polyhedral complex of the tropical equilibration branches.

The transition probabilities of the automaton are coarse grained properties of the statistical ensemble of trajectories for different initial conditions. Given a state and a minimal branch close to it, it will depend on the actual trajectory to which other branch the system will be close to next. However, when initial data and the full trajectory are not known, the automaton will provide estimates of where we go next and with which probability.
For the example studied, the branch B1 is a globally attractive sink: starting from anywhere, the automaton will reach B1 with probability one. This branch contains the unique stable steady state of the initial model. Figure~\ref{example2} bottom right shows the structure of most probable branches, the ones in which the systems spends most of his time. The branches B1, B3 and B2 correspond to different compositions of the membrane and of the endosome, rich in the receptor RI, rich in the receptor RII and rich in both types of receptors, respectively. Even if this
compositon is changed on wide domains of orders (planes in the space of orders), the concentrations
of effectors are robust (are more constained than the concentrations of receptors).

\section{Conclusion}
We have presented a method to coarse grain the dynamics of a smooth biochemical reaction network
to a discrete symbolic dynamics of a finite state automaton. The coarse graining was obtained
by two methods, approximated eigenvectors for mono-molecular networks and minimal branches
of tropical equilibrations for more general mass action nonlinear networks.
The two methods are compatible one to another, because when applied to monomolecular networks
the method based on tropical geometry detects all the transitions indicated by approximated
eigenvectors.
For both methods the automaton has a small number of states, less than the number of species
in the first method and the number of minimal tropical branches in the second method.
The coarse grained automaton can be used for studying statistical properties of biochemical
networks such as occurrence and stability of temporal patterns, recurrence, periodicity and
attainability  problems.
The coarse graining can be performed in a hierarchical way. For the nonlinear
example studied in the paper we computed only the full tropical equilibrations that
stand for the lowest order in the hierarchy (coarsest model). As discussed in Section~3
we can also consider partial equilibrations when slow variables are not equilibrated and thus refine
the automaton.
Our approach extends the notion of steady states of a network and propose a simple
recipe to characterize and detect metastable states. Most likely metastable states have biological
importance because the network spends most of its time in these states. The itinerancy of the network,
described as the possibility of transitions from one metastable state to another is paramount
to the way neural networks compute, retrieve and use information \cite{tsuda1991chaotic}
and can have similar
role in biochemical networks.

\paragraph{Acknowledgements}
O.R and A.N are supported by INCa/Plan Cancer grant N$^\circ$ASC14021FSA.


\newpage
\setcounter{page}{1}

\paragraph{Appendix 1: Proof of Proposition~\ref{eigenvectors}.}
Let us consider that $r_k^k=1$. Taking $r_k^j=0$
for all predecessors $j$ of $k$ and for all other
nodes that lead to $k$ by the flow $\Phi$
satisfy Eq.\eqref{recurrence1}(main body text) with  $\lambda = - \kappa_k$.
The same is valid for all the nodes that do not lead to $k$
and are not accessible from $k$. Remain the nodes that
are accessible from $k$. Let $j$ be such a node. Then
$j= \Phi^m(k)$ for some $m>0$.
 Eq.\eqref{recurrence1}(main body text) implies that
$$\kappa_{\Phi^{l-1}(k)} r_{\Phi^{l-1}(k)}^k =
(-\kappa_k + \kappa_{\Phi^{l}(k)}) r_{\Phi^{l}(k)}^k, \, \text{ for }
1 \leq l \leq m.$$
Thus
$r_{\Phi^{m}(k)}^k =
\frac{\kappa_k}{-\kappa_k + \kappa_{\Phi(k)}} \times
\frac{\kappa_{\Phi(k)}}{-\kappa_k + \kappa_{\Phi^2(k)}} \times
\ldots \times
\frac{\kappa_{\Phi^{m-1}(k)}}{-\kappa_k + \kappa_{\Phi^m(k)}}.$
Suppose that
$\kappa_k < \kappa_{\phi^{l}(k)}$ for $l=1,\ldots,m-1$
and  $\kappa_{\phi^{m}(k)} < \kappa_k$.
Using Lemma\ref{totalseplemma}(main body text) it follows $r_{\Phi^{m}(k)}^k = -1$. If any of the previous inequality does not
hold then at least one factor in the expression of
$r_{\Phi^{m}(k)}^k$ vanishes and the remaining factors are finite, thus
$r_{\Phi^{m}(k)}^k = 0$.
Consider now that $l_k^k=1$. Taking $l_k^j=0$
for all the nodes $j$ that can be obtained from
$k$ and for all other
nodes that do not lead to $k$ by the flow $\Phi$
satisfy Eq.\eqref{recurrence2}(main body text) with  $\lambda = - \kappa_k$.
The remaining nodes are all leading to $k$.  Let $j$ be such a node. Then
$k= \Phi^m(j)$ for some $m>0$.
 Eq.\eqref{recurrence2} (main body text)
implies that
$$\kappa_{\Phi^{l-1}(j)} l_{\Phi^{l}(j)}^k =
(-\kappa_k + \kappa_{\Phi^{l-1}(j)}) l_{\Phi^{l-1}(j)}^k, \, \text{ for }
1 \leq l \leq m.$$
Hence
$l_{j}^k =
\frac{\kappa_{j}}{-\kappa_k + \kappa_{j}} \times
\frac{\kappa_{\Phi(j)}}{-\kappa_k + \kappa_{\Phi(j)}} \times
\ldots \times
\frac{\kappa_{\Phi^{m-1}(j)} }{-\kappa_k + \kappa_{\Phi^{m-1}(j)}}.$
Suppose that
$\kappa_{\Phi^{l}(j) }  > \kappa_k$, for all $l = 0,\ldots,m-1$. Using Lemma\ref{totalseplemma}(main body text) it follows
$l_{j}^k = 1$. If one of these inequalities is not satisfied for a $l = 0,\ldots,m-1$
then the corresponding factor in the expression of $l_{j}^k$ vanishes and $l_{j}^k=0$.

The above formulas cover the zero eigenvalue case if
we consider that $\kappa_k=0$ for $k$ being the sink. It
follows that $r^0_k=1$ and
$r^0_j=0$ elsewhere. Furthermore, $l^0_j=1$ for all $j$.

\paragraph{Appendix 2: Algorithm for reduction of monomolecular networks with total separation
separation.}

This algorithm consists of three steps.

{\bf I. Constructing of an auxiliary reaction network: pruning.}

For each $A_i$ branching node (substrate of several reactions) let us define $\kappa_i$
as the maximal kinetic constant for reactions $A_i \to A_j$:
$\kappa_i =\max_j \{k_{ji}\}$. For correspondent $j$ we use the notation
$j = \phi(i)$: $\phi(i)={\rm arg \,max}_j \{k_{ji}\}$.

An auxiliary reaction network $\mathcal{V}$ is the set of reactions obtained by keeping
only $A_i \to
A_{\phi(i)}$ with kinetic constants $\kappa_i$ and discarding the other, slower reactions.
Auxiliary networks have no branching, but they can have cycles and confluences.
The correspondent
kinetic equation is
\begin{equation}\label{auxkinet}
\dot{c}_i =  -\kappa_i c_i + \sum_{\phi(j)=i} \kappa_j c_j,
\end{equation}

If the auxiliary network contains no cycles, the algorithm stops here.

{\bf II gluing cycles and restoring cycle exit reactions}

In general, the auxiliary network $\mathcal{V}$ has several
cycles $C_1,C_2, ...$ with periods $\tau_1, \tau_2,...>1$.

These cycles will be ``glued" into points and all nodes in the cycle $C_i$,
will be replaced by a single vertex $A^i$.
Also, some of the reactions that were pruned in the first part of
the algorithm are restored with renormalized rate constants. Indeed, reaction exiting a
cycle are needed to render the correct dynamics: without them, the total
mass of the cycle is conserved, with them the mass can also slowly leave the
cycle.
Reactions $A \to B$ exiting from cycles
($A \in C_i$, $B \notin C_i$)
are changed into  $A^i \to B$ with the rate
constant renormalization: let the cycle $C^i$ be the following
sequence of reactions $A_{1} \to A_{2} \to ... A_{\tau_i} \to
A_1$, and the reaction rate constant for $A_i \to A_{i+1}$ is
$k_i$ ($k_{\tau_i}$ for $A_{\tau_i} \to A_1$). For the limiting (slowest)
reaction of the cycle $C_i$ we use notation $k_{\lim \, i}$. If
$A=A_j$ and $k$ is the rate reaction for $A \to B$, then the new
reaction $A^i \to B$ has the rate constant $k k_{\lim \, i}/ k_j$.
This rate is obtained using quasi-stationary distribution for the cycle.
If kinetic constants are expressed as powers of a small
positive parameter $\epsilon$, i.e., if $k = \epsilon^\gamma$, then the order
of the constant has to be changed according to the rule $\gamma \to \gamma + \gamma_{lim} - \gamma_j$,
where $\gamma$, $\gamma_{lim \, i}$, $\gamma_j$ are the orders of the constants
$k$, $k_{\lim \, i}$ and  $k_j$, respectively.

The new auxiliary network $\mathcal{V}^1$ is computed for the network of glued cycles.
Then we decompose it into cycles, glue them, iterate until a acyclic
network is obtained $\mathcal{V}^n$.

{\bf III Restoring cycles}

The dynamics of species inside glued cycles is lost after the
second part.
A full multi-scale approximation (including relaxation
inside cycles) can be obtained by restoration of cycles.
This is done starting from the acyclic auxiliary
network $\mathcal{V}^n$ back to $\mathcal{V}^1$ through the
hierarchy of cycles. Each cycle is restored according to the
following procedure:

For each glued cycle node $A^m_{i}$, node of $\mathcal{V}^{m}$,

\begin{itemize}
\item Recall its nodes $A^{m-1}_{i1} \to A^{m-1}_{i2} \to ...
A^{m-1}_{i \tau_i} \to A^{m-1}_{i1}$; they form a cycle of length $\tau_i$.
\item Let us assume that the limiting step in $A^m_{i}$ is $A^{m-1}_{i
\tau_i} \to A^{m-1}_{i1}$
\item Remove $A^m_{i}$ from $\mathcal{V}^m$
\item Add $\tau_i$ vertices $A^{m-1}_{i1} ,
A^{m-1}_{i2}, ... A^{m-1}_{i\tau_i}$ to $\mathcal{V}^m$
\item Add to $\mathcal{V}^m$ reactions $A^{m-1}_{i1} \to A^{m-1}_{i2} \to ... A^{m-1}_{i
\tau_i}$ (that are the cycle reactions without the limiting step)
with correspondent constants from $\mathcal{V}^{m-1}$
\item If there exists an outgoing reaction $A^m_i \to B$ in
$\mathcal{V}^m$ then we substitute it by the reaction $A^{m-1}_{i
\tau_i} \to B$ with the same constant, i.e. outgoing reactions
$A^m_i \to...$ are reattached to the beginning of the limiting steps
\item If there exists an incoming reaction in the form $B \to
A^m_i$, find its prototype in $\mathcal{V}^{m-1}$ and restore it
in $\mathcal{V}^{m}$
\item If in the initial $\mathcal{V}^{m}$ there existed a ``between-cycles" reaction $A^m_i \to A^m_j$ then we find the
prototype in $\mathcal{V}^{m-1}$, $A \to B$, and substitute the
reaction by $A^{m-1}_{i \tau_i} \to B$ with the same constant, as
for $A^m_i \to A^m_j$ (again, the beginning of the arrow is reattached to the head of the limiting step in $A^{m}_{i}$)
\end{itemize}

\begin{figure}
\begin{center}
\scalebox{0.38}{
\begin{tikzpicture}
 \SetUpEdge[lw         = 0.5pt,
            color      = black,
            labelstyle = {sloped,scale=2}]
  \SetVertexMath
       \tikzset{VertexStyle/.style={scale=1.5,
       draw,
            shape = circle,
            line width = 1pt,
            color = black,
            outer sep=1pt}}
  \Vertex[x=0,y=6]{A1}
  \Vertex[x=4+1.2,y=6+2.8]{A2}
  \Vertex[x=8,y=6]{A3}
  \Vertex[x=0,y=0]{A4}
  \Vertex[x=4+1.2,y=2.8]{A5}
  \Vertex[x=8,y=0]{A6}
\tikzset{EdgeStyle/.style={->, line width = 5}}
\Edge[label=$1$](A1)(A2)
\tikzset{EdgeStyle/.style={post,line width = 2.5}}
\Edge[label=$6$](A2)(A3)
\tikzset{EdgeStyle/.style={post,line width = 3.5}}
\Edge[label=$4$](A3)(A1)
\tikzset{EdgeStyle/.style={post,line width = 1}}
\Edge[label=$9$](A4)(A5)
\tikzset{EdgeStyle/.style={post,line width = 3}}
\Edge[label=$5$](A5)(A6)
\tikzset{EdgeStyle/.style={post,line width = 4.5}}
\Edge[label=$2$](A6)(A4)
\tikzset{EdgeStyle/.style={post,line width = 4}}
\Edge[label=$3$](A1)(A4)
\tikzset{EdgeStyle/.style={post,line width = 2}}
\Edge[label=$7$](A2)(A5)
\tikzset{EdgeStyle/.style={post, thick, bend right = 10,line width = 0.5}}
\Edge[label=$10$](A3)(A6)
\tikzset{EdgeStyle/.style={post,bend right = 10,line width = 1.5}}
\Edge[label=$8$](A6)(A3)
\draw(4,-1.5)node[above,right, scale =3] {a)};

\begin{scope}[xshift=15cm]
  \Vertex[x=0,y=6]{A1}
  \Vertex[x=4+1.2,y=6+2.8]{A2}
  \Vertex[x=8,y=6]{A3}
  \Vertex[x=0,y=0]{A4}
  \Vertex[x=4+1.2,y=2.8]{A5}
  \Vertex[x=8,y=0]{A6}
\tikzset{EdgeStyle/.style={->, line width = 5}}
\Edge[label=$1$](A1)(A2)
\tikzset{EdgeStyle/.style={post,line width = 2.5}}
\Edge[label=$6$](A2)(A3)
\tikzset{EdgeStyle/.style={post,line width = 3.5}}
\Edge[label=$4$](A3)(A1)
\tikzset{EdgeStyle/.style={post,line width = 1}}
\Edge[label=$9$](A4)(A5)
\tikzset{EdgeStyle/.style={post,line width = 3}}
\Edge[label=$5$](A5)(A6)
\tikzset{EdgeStyle/.style={post,line width = 4.5}}
\Edge[label=$2$](A6)(A4)
\draw(4,-1.5)node[above,right, scale =3] {b)};
\end{scope}
\begin{scope}[xshift=-5cm,yshift=-12cm]
  \Vertex[x=4.4,y=6.933]{A1+A2+A3}
  \Vertex[x=4.4,y=0.9333]{A4+A5+A6}
\tikzset{EdgeStyle/.style={post, thick, bend right = 20, line width = 1}}
\Edge[label=$8$](A1+A2+A3)(A4+A5+A6)
\tikzset{EdgeStyle/.style={post, thick, bend right = 40, line width = 2}}
\Edge[label=$7$](A1+A2+A3)(A4+A5+A6)
\tikzset{EdgeStyle/.style={post, thick, bend right = 60, line width = 0.5}}
\Edge[label=$12$](A1+A2+A3)(A4+A5+A6)
\tikzset{EdgeStyle/.style={post,bend right = 20,line width = 0.5}}
\Edge[label=$15$](A4+A5+A6)(A1+A2+A3)
\draw(4,-1.5)node[above,right, scale =3] {c)};
\end{scope}
\begin{scope}[xshift=0cm,yshift=-12cm]
  \Vertex[x=4.4,y=6.933]{A1+A2+A3}
  \Vertex[x=4.4,y=0.9333]{A4+A5+A6}
\tikzset{EdgeStyle/.style={post, thick, bend right = 20, line width = 2}}
\Edge[label=$7$](A1+A2+A3)(A4+A5+A6)
\tikzset{EdgeStyle/.style={post,bend right = 20,line width = 0.5}}
\Edge[label=$15$](A4+A5+A6)(A1+A2+A3)
\draw(4,-1.5)node[above,right, scale =3] {d)};
\end{scope}
\begin{scope}[xshift=5cm,yshift=-12cm]
  \Vertex[x=4.4,y=6.933]{A1+A2+A3}
  \Vertex[x=4.4,y=0.9333]{A4+A5+A6}
\tikzset{EdgeStyle/.style={post, thick, bend right = 20, line width = 2}}
\Edge[label=$7$](A1+A2+A3)(A4+A5+A6)
\draw(4,-1.5)node[above,right, scale =3] {e)};
\end{scope}
\begin{scope}[xshift=15cm,yshift=-12cm]
\Vertex[x=0,y=6]{A1}
  \Vertex[x=4+1.2,y=6+2.8]{A2}
  \Vertex[x=8,y=6]{A3}
  \Vertex[x=0,y=0]{A4}
  \Vertex[x=4+1.2,y=2.8]{A5}
  \Vertex[x=8,y=0]{A6}
\tikzset{EdgeStyle/.style={->, line width = 5}}
\Edge[label=$1$](A1)(A2)
\tikzset{EdgeStyle/.style={post,line width = 3.5}}
\Edge[label=$4$](A3)(A1)
\tikzset{EdgeStyle/.style={post,line width = 3}}
\Edge[label=$5$](A5)(A6)
\tikzset{EdgeStyle/.style={post,line width = 4.5}}
\Edge[label=$2$](A6)(A4)
\tikzset{EdgeStyle/.style={post,line width = 2}}
\Edge[label=$7$](A2)(A4)
\draw(4,-1.5)node[above,right, scale =3] {f)};
\end{scope}
\end{tikzpicture}
}
\end{center}
\caption{
The successive steps of the reduction algorithm, illustrated for the prism model used in the paper.
a) is the initial model; b) is the auxiliary network resulting from step I, pruning; c) is the result of gluing 3 species cycles and renormalizing the exit reactions (the constants of orders
$3,7,10,8$ are renormalized to $3+6-1=8$,$7+6-6=7$, $10+6-4=12$, and $8+9-2=15$, respectively); d) is the auxiliary network after one more iteration; e) results from gluing and then restoring the 3 species cycles without the limiting step (constant of order 15); f) results from restoring the single species cycles without their limiting steps. }
\end{figure}
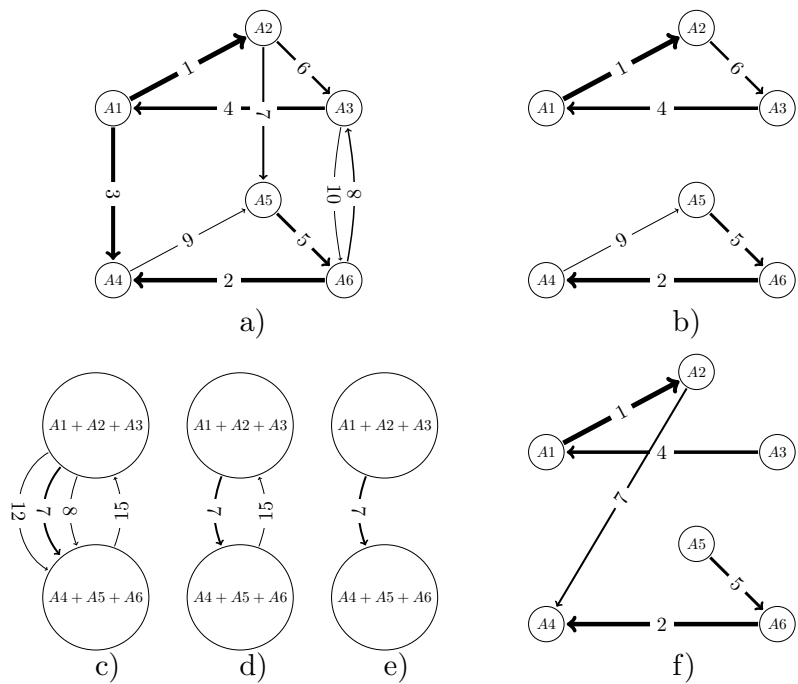

\paragraph{Appendix 3: Description of the TGFb model used in this paper.}

The model is described by the following system of differential equations
\begin{eqnarray}
\D{x_1}{t}&=& 	 k_{2} x_{2} - k_{1} x_{1} - k_{16} x_{1} x_{11}\notag \\
\D{x_2}{t}&= &	 	 k_{1} x_{1} - k_{2} x_{2} + k_{17} k_{34} x_{6}\notag \\
\D{x_3}{t}&= &	 	 k_{3} x_{4} - k_{3} x_{3} + k_{7} x_{7} + k_{33} k_{37} x_{18} -k_{6} x_{3} x_{5}\notag \\
\D{x_4}{t}&= &	 	 k_{3} x_{3} - k_{3} x_{4} + k_{9} x_{8} - k_{8} x_{4} x_{6}\notag \\
\D{x_5}{t}&= &	 	 k_{5} x_{6} - k_{4} x_{5} + k_{7} x_{7} + 2 k_{11} x_{9} - 2 k_{10} x_{5}^2 - k_{6} x_{3} x_{5} + k_{16} x_{1} x_{11}\notag \\
\D{x_6}{t}&= &	 	 k_{4} x_{5} - k_{5} x_{6} + k_{9} x_{8} + 2 k_{13} x_{10} - 2 k_{12} x_{6}^2 - k_{17} k_{34} x_{6} + k_{31} k_{36} x_{8} - k_{8} x_{4} x_{6}\notag \\
\D{x_7}{t}&= &	 	 k_{6} x_{3} x_{5} - x_{7} (k_{7} + k_{14})\notag \\
\D{x_8}{t}&= &	 	 k_{14} x_{7} - k_{9} x_{8} - k_{31} k_{36} x_{8} + k_{8} x_{4} x_{6}\notag \\
\D{x_9}{t}&= &	 	 k_{10} x_{5}^2 - x_{9} (k_{11} + k_{15})\notag \\
\D{x_{10}}{t}&=& 	 	 k_{15} x_{9} - k_{13} x_{10} + k_{12} x_{6}^2\notag \\
\D{x_{11}}{t}&= &	 	 k_{23} x_{14} - k_{30} x_{11}\notag \\
\D{x_{12}}{t}&= &	 	 k_{18} - x_{12} (k_{20} + k_{26}) + k_{30} x_{11} + k_{27} x_{15} - k_{22} k_{35} x_{12} x_{13}\notag \\
\D{x_{13}}{t}&= &	 	 k_{19} - x_{13} (k_{21} + k_{28}) + k_{30} x_{11} + k_{29} x_{16} - k_{22} k_{35} x_{12} x_{13}\notag \\
\D{x_{14}}{t}&= &	 	 k_{22} k_{35} x_{12} x_{13} - x_{14} (k_{23} + k_{24} + k_{25}) \notag \\
\D{x_{15}}{t}&= &	 	 k_{26} x_{12} - k_{27} x_{15}\notag \\
\D{x_{16}}{t}&= &	 	 k_{28} x_{13} - k_{29} x_{16}\notag \\
\D{x_{17}}{t}&= &	 	 k_{31} k_{36} x_{8} - k_{32} x_{17}\notag \\
\D{x_{18}}{t}&= &	 	 k_{32} x_{17} - k_{33} k_{37} x_{18}\notag
\end{eqnarray}

These variables are as follows:
\begin{itemize}
\item
Receptors on membrane: $x_{12}=$ RI, $x_{13}=$ RII, $x_{14}=$ LR.
\item
Receptors in the endosome: $x_{11}=$ LRe, $x_{15}=$ RIe, $x_{16}=$ RIIe.
\item
Transcription factors and complexes in cytosol: $x_1=$ S2c,  $x_3=$ S4c, $x_5=$ pS2c, $x_7=$ pS24c, $x_9=$ pS22c, $x_{18}=$ S4ubc.
\item
Transcription factors and complexes in the nucleus: $x_2=$ S2n, $x_4=$ S4n, $x_6=$ pS2n, $x_8=$ pS24n, $x_{10}=$ pS22n, $x_{17}=$ S4ubn.
\end{itemize}

\end{document}